\documentclass[a4]{iopart}
\usepackage[T1]{fontenc}
\usepackage[latin1]{inputenc}
\usepackage{geometry}
\usepackage{hyperref}
\usepackage{graphicx}
\usepackage{amssymb}

\makeatletter

\date{\today}
\def\be{\begin{equation}}
\def\bea{\begin{eqnarray}}
\def\eea{\end{eqnarray}}
\def\ee{\end{equation}}
\def\no{\nonumber}
\def \a{\alpha}

\begin{document}
\title{\bf {Fundamentals of the LISA Stable Flight Formation}}
\author{S. V. Dhurandhar$^1$, K. Rajesh Nayak$^{2,3}$, S. Koshti$^1$ and J-Y. Vinet$^2$}
 
\address{ $^1$IUCAA, Postbag 4, Ganeshkind, Pune - 411 007, India. 
\\ $^2$ARTEMIS, Observatoire de la Cote d'Azur, 
\\ BP 4229, 06304 Nice, France.
\\ $^3$ Present Address: The University of Texas at Brownsville, 
\\80 Fort Brown,
Brownsville, TX 78520, USA.
}

\begin{abstract}
The joint NASA-ESA mission LISA relies crucially on the stability of the three spacecraft  constellation. Each of the spacecraft is in heliocentric orbits forming a stable triangle. 
The principles of such a formation flight have been formulated long ago and analysis performed, but seldom presented if ever, even to LISA scientists. We nevertheless need these details in 
order to carry out theoretical studies on the optical links, simulators etc. In this article, 
we present in brief, a model of the LISA constellation, which we believe will be useful for the LISA community.
\end{abstract}

\maketitle
\section{Introduction}
LISA is a ESA-NASA mission for observing low frequency gravitational waves in
the frequency range from $10^{-5}$ Hz to 1 Hz \cite{RIP}. 
In order for LISA to operate
successfully, it is crucial that the three spacecraft which form the hubs of
the laser interferometer in space maintain nearly constant distances between
them, though their order of magnitude is $5\times 10^6$ km. 
The existence of orbits having this property was firstly reported by
Bender \cite{PB} as the basis of LISA. In order to thoroughly study the
optical links and light propagation between these moving stations, we however need
a detailed model of the LISA configuration. We therefore find it useful to recall
explicitly the not so trivial principles of a stable formation flight.
In this brief work, we firstly study three Keplerian orbits around the Sun 
with small eccentricities and adjust the orbital parameters so that the spacecraft 
form an equilateral triangle with nearly constant distances between them. Then we find
that to the first order in the parameter $\a = l/2R$, where 
$l \sim 5 \times 10^6$ km, is the distance between two spacecraft and $R = 1$ A. U. 
$\sim 1.5 \times 10^8$ km, the distances between spacecraft are exactly constant; 
any variation in arm-lengths should result from higher orders in $\a$ or from external 
perturbations of Jupiter and the secular effect due to the Earth's gravitational field. 
(The eccentricity $e$ is related in a simple way to $\a$ and is proportional to $\a$ to 
the first order in $\a$.) In fact our analysis shows that such formations are possible
with any number of spacecraft provided
they lie in a magic plane making an angle of $60^{\circ}$ with the ecliptic. We establish 
this general result with the help of the Hill's or Clohessy-Wiltshire (CW) equations \cite{CW}. 
\section{The choice of orbits}
\subsection{The exact orbits}
The exact orbits of the three spacecraft are constructed so that to the first order in 
the parameter $\a$, the distances between any two spacecraft remain constant. Below we 
give such a choice of orbits. This choice is clearly not unique and other choices are 
possible which satisfy some criteria of optimality such as the distances between 
spacecraft vary as little as possible.
\par
We construct the orbit of the first spacecraft and then obtain the other
two orbits by rotations of $120^{\circ}$ and $240^{\circ}$. The equation of an elliptical 
orbit in the $(X-Y)$ plane is given by \cite{baker},
\be
X = R(\cos \psi + e), ~~~ Y = R \sqrt{1 - e^2} \sin \psi,
\label{orbit}
\ee
where $R$ is the semi-major axis of the ellipse, $e$ the eccentricity and $\psi$
the eccentric anomaly. The focus is at the origin. The eccentric anomaly is 
related to the mean anomaly $\Omega t$ by,
\be
\psi + e \sin \psi = \Omega t,
\label{anml}
\ee
where $t$ is the time and $\Omega$ the average angular velocity. We have chosen the zero 
of time when the particle is at the
farthest point from the focus (this is contrary to what most books do and because of this 
choice of initial condition we have a positive sign instead of a negative sign on the 
left hand side of Eq.(\ref{anml})\ ). 
\par
We choose the barycentric frame with coordinates $(X, Y, Z)$ as follows: The ecliptic plane 
is the $X-Y$ plane and we first consider a circular reference orbit of radius 1 A. U. 
centered at the Sun. The plane of the LISA triangle makes an angle of $60^{\circ}$ with 
the ecliptic plane. As we shall see later, we deduce from the CW equations that this allows 
constant inter-spacecraft distances to the first order in $\a$. This fact dictates the choice 
of orbits of the spacecraft formation. We
choose spacecraft 1 to be at its highest point (maximum Z) at $t = 0$. This
means that at this point, $\psi = 0$ and $Y = 0$. Thus to obtain the orbit of the first
space-craft we must rotate the orbit in Eq. (\ref{orbit}) by a small angle 
$\epsilon$ about the $Y-$axis so that the spacecraft 1 is lifted by an appropriate distance above
the $X-Y$ plane. In order to obtain the inclination of $60^{\circ}$, the spacecraft must have its 
$Z$-coordinate equal to $l/2$. The geometry of the configuration is shown in Figure 1. 
\begin{figure}[h]
		\begin{center}
\includegraphics[width=0.8\textwidth]{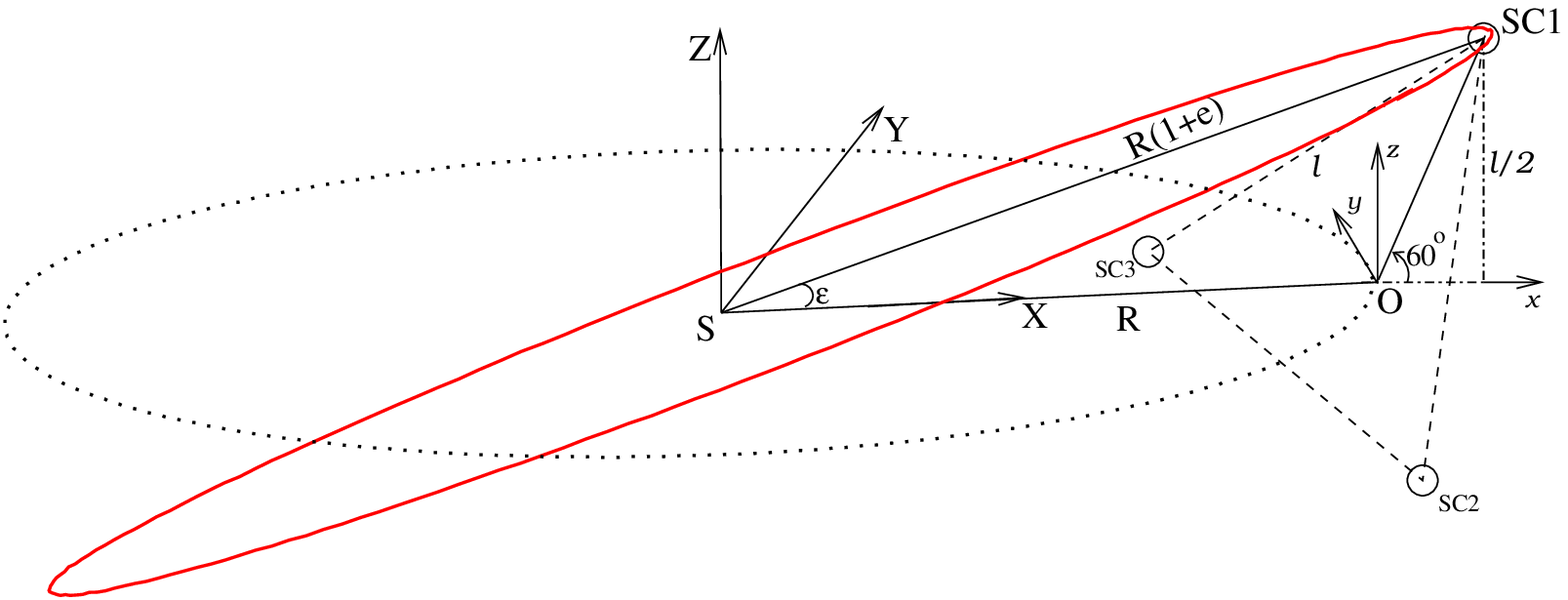}
\caption{The figure shows the geometry of the orbits and of LISA. The barycentric frame 
is labelled by $(X, Y, Z)$ while the CW frame is labelled by $(x, y, z)$. SC1, SC2 and 
SC3 denote the three spacecraft. The radius of the reference orbit is taken to be $R = 1$ 
A. U. and S denotes the Sun.}
\label{geom}
\end{center}
\end{figure}  
From the geometry $\epsilon$ and $e$ are obtained as,
\bea
\tan \epsilon &=& \frac{\a}{1 + \a /\sqrt{3}} \, ,\\ 
e &=& \left( 1 + \frac{2}{\sqrt{3}}\a + \frac{4}{3} \a^2 \right)^{1/2} - 1 \, ,
\eea
\par
and the orbit equations for the spacecraft 1 are given by:
\bea
X_1 &=& R(\cos \psi_1 + e) \cos \epsilon, \no \\
Y_1 &=& R \sqrt{1 - e^2} \sin \psi_1, \no \\
Z_1 &=& R(\cos \psi_1 + e) \sin \epsilon. 
\label{tltorb}
\eea  
The eccentric anomaly $\psi_1$ is implicitly given in terms of $t$ by,
\be
\psi_1 + e \sin \psi_1 = \Omega t.
\ee
The orbits of the spacecraft 2 and 3 are obtained by rotating the orbit 
of spacecraft 1 by 
$2 \pi / 3$ and $4 \pi/3$ about the $Z-$axis; the phases $\psi_2, \psi_3$ 
however, must be adjusted so that the spacecraft are about the distance $l$ 
from each other. The orbit equations of spacecraft $k = 2, 3$ are:
\bea
X_k &=&  X_1 \cos \left[ \frac{2 \pi}{3}(k-1) \right] - Y_1 \sin \left[ \frac{2 \pi}{3}(k-1) \right] \, , \no \\
Y_k &=&  X_1 \sin \left[ \frac{2 \pi}{3}(k-1) \right] +  Y_1 \cos \left[ \frac{2 \pi}{3}(k-1) \right] \, , \no \\
Z_k &=& Z_1 \, ,
\label{orbits}
\eea
with the caveat that the $\psi_1$ is replaced by the phases $\psi_k$ where they are implicitly given by,
\be
\psi_k + e \sin \psi_k = \Omega t - (k -1) \frac{2 \pi}{3}.
\ee 
These are the exact equations of the orbits of the three spacecraft. With these orbits 
the inter-spacecraft distance varies upto about 100,000 km. In Figure 2 we show how the 
inter-spacecraft distances vary over the course of a year. 
\begin{figure}[h]
\centering
\includegraphics[width=0.6\textwidth]{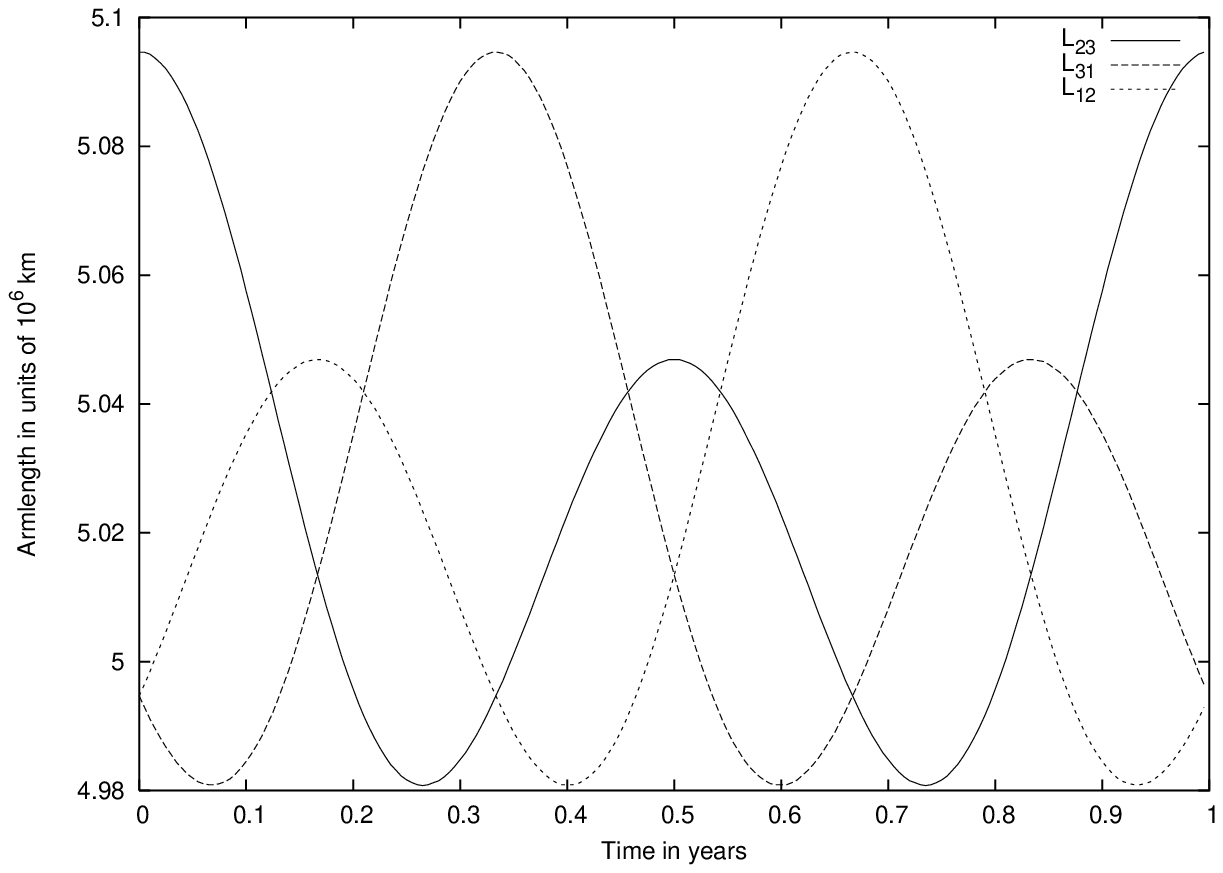}
\caption{The variation of the lengths of the arms of LISA (the breathing modes) 
is shown in units of $10^6$ km, when the exact orbits are computed. To the first 
order in $\a$ the lengths of the arms remain constant and are equal to $l = 5 \times 10^6$ km.}
\label{brth}
\end{figure} 
Note that there are other choices of $e$ and $\epsilon$ close to the above values for the 
three orbits which give smaller variations in the armlengths.
\subsection{The orbits to first order in $\a$ }
In this subsection we obtain the orbits to the first order in $\alpha$. The 
tilt $\epsilon$ and the eccentricity $e$ are given to this order by,
\be
\tan \epsilon = \a, ~~~~~~~~e = \frac{\a}{\sqrt{3}}.
\ee
We find that $e$ is proportional to $\a$ and, $\epsilon \sim 1.7 \times 10^{-2}$ and 
$e \sim 10^{-2}$. Then to this order and now writing $\a$ in terms of $e$, Eqs.(\ref{tltorb}) become: 
\bea
X_1 &=& R(\cos \psi_1 + e), \no \\
Y_1 &=& R \sin \psi_1, \no \\
Z_1 &=& \sqrt{3} e R \cos \psi_1 , 
\eea  
where the eccentric anomaly can be explicitly solved for to the first order in $e$ in terms of the time $t$:  
\be
\psi_1 = \Omega t - e \sin \Omega t.
\ee
The approximate orbits of the spacecraft 2 and 3 can be obtained, as before, by 
rotating the orbit of spacecraft 1 by $2 \pi/3$ and $4 \pi/3$ respectively about 
the $Z$-axis as in Eq.(\ref{orbits}). The corresponding phases $\psi_2$ and $\psi_3$ 
now, can be explicitly obtained in terms of $t$:
\be
\psi_k = \Omega t - (k - 1) \frac{2 \pi}{3} - e \sin \left[\Omega t - (k - 1) \frac{2 \pi}{3}\right].
\ee 
In the next section we prove that to the first order in $\a$, the distance between any two
space-craft is $l$, that it is a constant and remains so {\it at all times}; 
{\it the LISA constellation moves rigidly as an equilateral triangle with its centroid 
tracing a circle with radius of 1 A. U. with the Sun as its centre.} To check this from 
the above equations is straightforward. We can compute the distance between spacecraft 
1 and 2, which at the lowest order in $\a$ proves to be $2 \a R$, the two other distances 
are equal to the preceding
by symmetry. This model succeeded because we already knew the result. In the next section, 
we show with the help of a more sophisticated formalism, how this case is a special case of
a much more general result and that stable formations with infinite number of spacecraft 
are possible. 
This result is important because we then have large number of flight formations to choose 
from. Depending on the required physical criteria optimal flight formations may be selected. 
\section{The CW frame}
Clohessy and Wiltshire \cite{CW} make a transformation to a frame - the CW frame $(x, y, z)$ 
which has its origin on the reference orbit and also rotates with angular velocity $\Omega$.  
The $x$ direction is normal and coplanar with the orbit, the $y$ direction is
tangential and comoving, and the $z$ direction is chosen orthogonal to the orbital plane.  
They write down the linearised dynamical equations for test-particles in the 
neighbourhood of a reference particle (such as the Earth). Since the
frame is noninertial, Coriolis and centrifugal forces appear in addition to the tidal forces. 
The CW equations for a free test particle of coordinates $(x,y,z)$ are:
\begin{equation}
\left\{
\begin{array}{l}
\ddot{x}-2\Omega \dot{y} -3\Omega^2 x\ =\ 0 \, ,  \\
\ddot{y} +2\Omega \dot{x} \ = \ 0 \, , \\
\ddot{z}+\Omega^2 z \ = \ 0 \, ,
\end{array}
\right.
\label{clowi}
\end{equation}
where $\Omega \equiv 2\pi/1$ year.
The general solution, depending on six arbitrary parameters is:
\begin{eqnarray}
\left\{
\begin{array}{l}
x(t)\ = \ \frac{\dot{x}_0}{\Omega}\,\sin \Omega t 
-\left( 3x_0+\frac{2\dot{y}_0}{\Omega}\right)\,\cos \Omega t 
+2\left( 2x_0+\frac{\dot{y}_0}{\Omega}\right) \, , \\
y(t)\ = \ \left( 6x_0+\frac{4\dot{y}_0}{\Omega}\right) \,\sin \Omega t 
+\frac{2\dot{x}_0}{\Omega}\,\cos \Omega t  -3(2\Omega x_0 + \dot{y}_0)t+
\left(y_0-\frac{2\dot{x}_0}{\Omega} \right) \, ,\\
z(t)\ = \ z_0 \cos \Omega t + \frac{\dot{z}_0}{\Omega} \, \sin  \Omega t \, . 
\end{array}
\right.
\label{clowisol}
\end{eqnarray}
We observe that both $x$ and $y$ contain constant terms and $y$ also contains a 
term linear in $t$. The constant term in $y$ is merely  an offset and could be removed 
without loss of generality by a trivial translation of coordinate along the same orbit. 
The removal of the $x$ offset also removes the linear term in $t$ 
(the runaway solution). In contrast with the $y$ offset, the $x$ 
offset corresponds to a  different orbit with a different period 
than that of the reference  particle, namely, the origin of the 
CW frame. Thus the only actual and 
important requirement is that of vanishing of the $x$ offset term. This  
term represents Coriolis acceleration in the $y$ direction and comes from 
integrating  the $y$ equation in the CW equations (\ref{clowi}). 
If we require a solution with no offsets, we must have:
\bea
\dot{y}_0+2\Omega x_0=0, \no \\
\dot{x}_0-\frac{1}{2}\Omega y_0=0.
\label{rnwy}
\eea
With these additional constraints on the initial conditions, the bounded and centred solution is:
\bea
x(t)=\frac{1}{2} y_0 \sin \Omega t + x_0 \cos \Omega t ,
\no \\
y(t)=y_0 \cos \Omega t  - 2x_0  \sin \Omega t ,
\no \\
z(t) = z_0 \cos \Omega t + \frac{\dot{z}_0}{\Omega}  \sin \Omega t.
\eea
If moreover we require the distance of the particle from the origin
to be constant, equal to $d$, say, we get the following equation:
\bea
\left(
\frac{1}{4} y_0^2 + 4 x_0^2 + \frac{\dot{z}_0^2}{\Omega^2}
\right) \sin^2 \Omega t  +
(x_0^2+y_0^2+z_0^2) \cos^2 \Omega t & + & \no \\
\left( \frac{2z_0 \dot{z}_0}{\Omega}-3 x_0 y_0 \right) \sin \Omega t  \cos \Omega t 
 =  d^2. & \ & \
\eea
After identifying the terms of frequencies $0$ and $2\Omega$ (sin and cos), we
obtain the two equations:
\bea
z_0^2 -  \frac{\dot{z}_0^2}{\Omega^2} &=& 3\left( x_0^2 - \frac{1}{4} y_0^2 \right) \, , \no \\
\frac{2z_0 \dot{z}_0}{\Omega} &=& 3 x_0 y_0 \, .
\eea
Adding the first to $i$ times the second yields the complex condition:
\be
\left(z_0+ i \frac{\dot{z}_0}{\Omega} \right)^2=3\left(x_0 + i \frac{y_0}{2} \right)^2 \, ,
\ee
from which we obtain,
\be
z_0 \ = \ \mu \sqrt{3} x_0 \ \ , \ \ \frac{\dot{z}_0}{\Omega} \ = \
\frac{1}{2} \mu \sqrt{3} y_0,
\ee
where $\mu = \pm 1$. The solutions satisfying the requirements of 
(i) no offset and (ii) fixed distance to origin are finally of the form,
\begin{equation}
\left\{
\begin{array}{l}
x(t) \ = \ \frac{1}{2}\,\rho_0 \cos (\Omega t - \phi_0) \\
y(t) \ = \ -\rho_0 \sin (\Omega t - \phi_0) \\
z(t) \ = \  \mu \rho_0 \frac{\sqrt{3}}{2}\, \cos (\Omega t - \phi_0),
\end{array}
\right.
\label{CWP}
\end{equation}
where
\be
\rho_0 \ = \ \sqrt{4 x_0^2 + y_0^2}, \ \
 \ \  \tan \phi_0 \ =\ \frac {y_0} {2 x_0} \, .
\ee
The initial conditions are now expressed in terms of $(\rho_0, \phi_0)$ instead of $(x_0,y_0)$.
We call the solutions satisfying the above requirements as {\it stable}. 
The results that we obtained by taking Keplerian orbits to the first order in $\a$, are the same as 
those obtained by using the preceding CW equations. In the CW frame the equations 
of the orbits simplify and it is easy to verify the result. The transformation is only 
in the $(x,y)$ plane; the $z$ coordinate is undisturbed. Since we have chosen the 
reference orbit to be the circle centred at the Sun and radius of $R = 1$ A. U., 
the CW frame $(x, y, z)$ is related to our barycentric $(X, Y, Z)$ frame by:
\bea
x &=& (X - R \cos \Omega t) \cos \Omega t + (Y - R \sin \Omega t) \sin \Omega t, \no  \\
y &=& -(X - R \cos \Omega t) \sin \Omega t + (Y - R \sin \Omega t) \cos \Omega t, \no \\ 
z &=& Z. 
\eea
The orbit equations for the three spacecraft derived in the last section, now simplify 
and can again be written in a compact form:
\bea
x_k &=& e R \cos \left[ \Omega t - (k - 1) \frac{2 \pi}{3}\right] , \no\\
y_k &=& -2 e R \sin \left[ \Omega t - (k - 1) \frac{2 \pi}{3}\right], \no\\
z_k &=& \sqrt{3} e R \cos \left[ \Omega t - (k - 1) \frac{2 \pi}{3} \right],
\eea
where $k = 1, 2, 3$ labels the three space-craft. One immediately recognizes
the form of Eqs.(\ref{CWP}) for the special case of $\mu=1$ with the initial conditions 
$\rho_0 = 2 e R$ and $\phi_0 = 2 \pi (k - 1)/3 $. The symmetry is now obvious. 
It is straightforward to verify that the distance between any two spacecraft is
$l$. Thus the LISA spacecraft constellation rigidly moves as an equilateral triangle 
of side $l$ in this approximation.
\par
In fact, it is possible to establish a general result: {\it In the CW frame there are
just two planes which make angles of $\pm \pi/3$ with the (x-y) plane, in which test
particles obeying CW equations and the stability conditions (as defined above),
perform rigid rotations about the origin with angular velocity $- \Omega$.}   
\par
To see this, consider a test particle at arbitrary $(\rho_0, \phi_0)$ whose 
orbit is parametrized  by Eqs.(\ref{CWP}). Consider the frame which is obtained 
from the CW frame 
$(x, y, z)$, by first rotating about the $y$-axis by $\mu \pi /3$ to obtain the 
intermediate frame $(x', y', z')$ and then rotating this frame about the $z'$-axis 
by $-\Omega t$. The first rotation transforms the particle trajectories to lie in
the $(x', y')$ plane. The second rotation by $-\Omega t$ about the $z'$-axis 
makes the particle in this new frame $(x'', y'', z'')$ {\it stationary}! 
Thus we have in the double-primed coordinates:
\be
x''(t) \ = \   \rho_0 \cos \phi_0,  \ \ y''(t) \ = \   \rho_0 \sin \phi_0 \, ,
\ee
showing that the particle is at rest in the new rotating frame. There is thus 
a one to one mapping from the set of all stable (as defined above) solutions 
of the CW equations to the two planes whose normals $\vec n $ are inclined at 
$30^{\circ}$ or $150^{\circ}$ with respect to the $x$ direction and rotating
at the angular velocity  $- \Omega$, the rotation axis being $\vec n$. The LISA
plane corresponds to the choice of $150^{\circ}$, and it is now clear that any 
particle at rest in this plane, remains at rest in it, so that any number
of spacecraft in this plane would remain at constant relative distances, at least
in the CW approximation, equivalent to a first order calculation in
the eccentricities. This further implies that so far as `rigid' flight formations
are desired, equilateral triangle is not the only choice. Arbitrary formations 
with any number of spacecraft are possible as long as they obey the CW equations 
and satisfy the stability requirements as detailed above. 
\section{Conclusion}
We have explicitly constructed three heliocentric spacecraft orbits which to 
the first order in eccentricities maintain equal distances between them which
is taken to be 5 million km. We have shown with the help of a more sophisticated
formalism - the CW equations - that there are two planes in the CW frame, in which
particles obeying the CW equations and satisfying stability requirements, namely,
no offsets (and hence no runaway behaviour) and maintaining equal distance from the 
origin, maintain their relative distances in the CW approximation which is equivalent
to a first order calculation in the eccentricities. This has the implication that 
formations not necessarily triangular and with any number of spacecraft are possible
as long as they obey the stability constraints and lie in any one of these planes;
their relative distances will be maintained within the CW approximation. This result
opens up new possibilities of spacecraft constellations with various geometrical 
configurations and any number of spacecraft which would be useful to future space missions. 
\ack
SD would like to thank IFCPAR for travel support and the Observatoire de la C\^ote D'Azur, 
Nice, France for local hospitality, where substantial part of this work was carried out. 
SK would like to acknowledge DST, India for the WOS-A. All the authors would like to 
thank E. Chassande-Mottin for detailed discussions.    
\vspace{24pt} 
%
%
%
%

\end{document}